\documentclass[12pt,preprint]{aastex}

\newcommand{\HI} {\ion{H}{1}}
\newcommand{\HII} {\ion{H}{2}}
\newcommand{\OI} {\ion{O}{1}}
\newcommand{\OII} {\ion{O}{2}}
\newcommand{\OIII} {\ion{O}{3}}
\newcommand{\NI} {\ion{N}{1}}
\newcommand{\NII} {\ion{N}{2}}
\newcommand{\SII} {\ion{S}{2}}
\newcommand{\SiII} {\ion{Si}{2}}

\def\ltsima{$\; \buildrel < \over \sim \;$}
\def\simlt{\lower.5ex\hbox{\ltsima}}
\def\gtsima{$\; \buildrel > \over \sim \;$}
\def\simgt{\lower.5ex\hbox{\gtsima}}

\slugcomment{REVISED VERSION, ApJ}

\begin{document}


\title{The \ion{H}{2} Regions of the Damped Lyman~$\alpha$ Absorber SBS~1543+593\footnotemark[1]}


\author{Regina E. Schulte-Ladbeck, Sandhya M. Rao, Igor O. Drozdovsky\footnotemark[2], 
David A. Turnshek, Daniel B. Nestor}
\affil{Department of Physics \& Astronomy, University of Pittsburgh, 3941 O'Hara St., Pittsburgh, PA 15260, USA}

\and

\author{Max Pettini}
\affil{Institute of Astronomy, Madingley Road, Cambridge, CB3 0HA, UK}


\footnotetext[1]{Based on observations made with the NASA/ESA Hubble 
Space Telescope obtained from the 
Space Telescope Science Institute, which is operated by the Association of 
Universities for Research in Astronomy, Inc., under NASA contract NAS 5-26555.}

\footnotetext[2]{Astronomical Institute, St. Petersburg State University,
                            198904 St. Petersburg, Russia}

\begin{abstract}

We report new imaging and spectroscopic observations of the damped
Ly$\alpha$ (DLA) galaxy SBS~1543+593, a nearby dwarf galaxy whose 
stellar disk is intersected by the sightline to the bright 
background QSO HS~1543+5921. 
{\it Hubble Space Telescope} imaging observations with WFPC2
in the F450W and F702W bands are used to measure the DLA galaxy's
properties and compile a catalog of its (candidate) \ion{H}{2}
regions.  Ground-based long-slit spectroscopy of the brightest
\ion{H}{2} region in the galaxy yields estimates of the star
formation rate (SFR) and of chemical abundances in the
galaxy's interstellar medium.
We find that SBS~1543+593 exhibits a 
SFR $\approx 0.006\,h^{-2}_{70} M_{\odot}$~yr$^{-1}$, 
or a SFR per unit area of 
$\approx~1.4 \times 10^{-4}\,h^{-2}_{70} M_\odot$~yr$^{-1}$~kpc$^{-2}$. 
We derive gas-phase abundances in the ionized gas of 
$12 + \log {\rm (O/H)} =  8.2 \pm 0.2$, which is about 1/3 of the solar value, and
$\log {\rm (N/O)} = -1.40^{+0.2}_{-0.3}$. These values are consistent with 
the morphologial appearance of SBS~1543+593, an 
Sm dwarf of $M_B -5 \log h_{70} = -16.8\pm0.2$ 
and of intermediate surface brightness. 
SBS~1543+593 is the first {\it bona fide} DLA for which abundances 
have been measured using emission-line diagnostics. 
When compared with future, high-resolution, ultraviolet spectroscopy, 
our results should prove key for interpreting 
abundance determinations in high redshift DLAs.

\end{abstract}


\section{Introduction}

Originally classified as a Seyfert galaxy, SBS~1543+593 
was in fact shown by Reimers \& Hagen (1998) to be the chance
superposition of a nearby ($z = 0.0096$) galaxy and a background
($z_{\rm em} = 0.807$) QSO, HS~1543+5921, 
discovered independently in the course
of the Hamburg Quasar Survey.  The QSO
sightline intersects the galaxy only 2.4\,arcsec 
($ = 0.5 h_{70}^{-1}$\,kpc) from 
its center.\footnote{Throughout this paper we
use $h_{70} = H_0/70$\,km~s$^{-1}$~Mpc$^{-1}$, where $H_{0}$
is the Hubble constant.} 

More recently, this interesting galaxy-QSO pair has been
studied in detail by Bowen, Tripp, \& Jenkins (2001a) and
Bowen et al. (2001b). Ultraviolet spectroscopy with the
{\it Hubble Space Telescope} ({\it HST\/}) revealed that
the galaxy produces a damped Ly$\alpha$ (DLA) 
absorption system in the spectrum of the QSO,
with a neutral hydrogen column density 
$\log N$(\HI)$= 20.35$ [where $N$(\HI) 
is in units of cm$^{-2}$].  From ground-based imaging, Bowen et al. (2001a) measured
a central surface brightness $\mu_R (0) = 22.6$\,mag~arcsec$^{-2}$
and an exponential scale-length $a = 7$\,arcsec ($1.4 h_{70}$\,kpc);
integrating over the surface brightness profile they deduced a
total magnitude $R = 16.3$, which in turn
corresponds to an absolute magnitude 
$M_R - 5 \log h_{70} = -16.7$. 
Radio observations of the 21\,cm emission line (Bowen et al. 2001b)
yielded a systemic heliocentric velocity 
$v_{\rm helio} = 2868$\,km~s$^{-1}$ and an \HI\ mass
$M_{\rm H~I}  = 1.5 \times 10^9 h_{70}^{-2} M_{\odot}$;
the dynamical mass is more than twice the neutral gas mass. 
From all of these properties, Bowen et al. concluded 
that SBS~1543+593 is low surface brightness (LSB)
dwarf spiral galaxy.

The idea that dwarf and LSB galaxies are responsible for at least
some DLAs was originally proposed by York et al. (1996) and 
subsequently re-examined by several authors
(e.g. Jimenez, Bowen, \& Matteucci 1999; 
Boissier, P{\' e}roux, \& Pettini 2002).
Observationally, this has now been confirmed in
a number of cases, at least at redshifts $z \simlt 1$.
In particular, Rao \& Turnshek (2000) have successfully identified many
low-redshift DLA systems, and are in the process of studying them
with direct imaging (e.g., Turnshek et al. 2001; Nestor et al. 2002;
Rao et al. 2003 and references therein). 
Indeed, many turn out to be dwarf and/or LSB galaxies. 
The combination of low absorber redshift, small impact parameter, 
and brightness of the background QSO ($B = 16.8$, Reimers \& Hagen 1998) 
makes SBS~1543+539 a highly promising system for investigating
the all-important DLA-galaxy connection in greater detail
than is possible in more distant examples.\footnote{
To our knowledge, SBS~1543+539 is the second lowest DLA galaxy cataloged
up to now. The closest example is the NGC~4203 
($v_{\rm helio} = 1117$\,km~s$^{-1}$)--Ton~1480 ($z_{\rm em} = 0.614$)
pair (Miller, Knezek, \& Bregman 1999), 
but the QSO in this case is $\sim 1$\,mag fainter than 
HS~1543+5921 in the $B$-band. Furthermore, the QSO-galaxy
impact parameter is $8.6 h_{70}^{-1}$\,kpc, 
nearly 20 times greater than that of
SBS~1543+539--HS~1543+5921.
Of course, any galaxy which presents a cross-section in \HI\ absorption
greater than $N$(\HI)\,$= 2 \times 10^{20}$~cm$^{-2}$
(the historical limit defining DLAs)
is capable of being a DLA galaxy. However, a background 
source is required to produce a continuum against which this \HI\ 
column can be measured.}

More generally, SBS~1543+539 offers one of the best opportunities
to compare element abundances measured from emission lines
in \HII\ regions with those from cool interstellar gas producing 
absorption lines. This is a very important consistency check,
as we now explain. Essentially all available information on
chemical abundances at high redshifts is from QSO absorption
line studies---data from other techniques are still very few 
and of limited accuracy (Pettini 2003).
Conversely, most of what we know about the chemical composition
of galaxies in the nearby universe is based on measurements
of nebular lines from star-forming regions (e.g. Garnett 2003).
While in the Milky Way and the Magellanic Clouds these two
different lines of enquiry give consistent results, there has
so far been very little opportunity to compare them in other
environments, at either high or low redshift.
Relevant to this point is the suggestion by Kunth \& Sargent (1986)
that the \HII\ regions of metal-poor galaxies
may be self-polluted with metals produced in the
immediate past and not yet mixed in the general interstellar 
medium (ISM). 
One such example {\it may} be the Blue Compact Dwarf (BCD) galaxy
I~Zw~18, but opinions are still divided as to whether the O/H 
ratio measured from interstellar absorption lines is lower than
the value found in the \HII\ regions (Kunth et al. 1994;
Pettini \& Lipman 1995; Aloisi et al. 2003; Lecavalier des Etangs et al. 2003).
In the $z = 2.7276$ Lyman break galaxy MS 1512-cB58, 
Pettini et al. (2002) found good agreement between 
the oxygen abundance measured from nebular emission 
lines and the interstellar abundances of other $\alpha$-capture 
elements, although nitrogen was apparently discrepant.

The SBS~1543+539--HS~1543+5921 pairing gives us the means to
address these issues with a modest amount of observational effort
and with fewer complications than I~Zw~18, where the interstellar
absorption lines are viewed against the (spatially extended)
background of a stellar cluster which is 
located {\it within} the galaxy, rather than a point source
behind it.
In anticipation of forthcoming
ultraviolet spectroscopy of HS~1543+5921 
with sufficient resolution for abundance determinations,
we present new imaging and spectroscopic observations
of SBS~1543+539 aimed at obtaining 
additional information about its properties 
not already covered in the studies by Bowen et al. (2001a,b).
In particular, we report the first measurements of O and N abundances 
in one of the \HII\ regions of the galaxy.
The outline of the paper is as follows. 
In \S2 we describe the
observations which consist of {\it HST} WFPC2 imaging
and ground-based optical spectroscopy.  
In \S3 we present our main results: 
the galaxy's \ion{H}{2} region luminosity function (\S3.1), its
star formation rate (\S3.2) and element abundances
(\S3.3).  These are compared with similar results for nearby 
dwarf galaxies and high redshift DLAs in \S4.
In \S5 we draw attention to specific problems which can be addressed
by comparing the data presented
here for SBS~1543+539 with those which will be obtained from
absorption line studies of its neutral interstellar medium.
Finally, we summarize our main conclusions in \S6.

\section{Observations and Data Processing}

\subsection{Imaging}

We imaged the field of SBS~1543+593 with the 
Wide Field and Planetary Camera (WFPC2) on {\it HST}
on 20 September 2000 as part of our GO program 8570.
The galaxy was located on the WF3 chip.
Exposures were obtained through the following filters:
F225W (4000\,s), F336W (3000\,s), F450W (1000\,s), and
F702W (1000\,s). Figure~1 is a greyscale 
reproduction\footnote{A colour version of this figure
is available in the electronic version of the paper.}
of an image formed by combining data from all four filters with the 
following color coding: red=F702W,
green=F450W, and blue=F336W+F255W.
As well as the QSO HS~1543+5921, a  foreground star
(to the north-west of the QSO), and several background galaxies,
the image shows the faint spiral arms of SBS~1543+593
traced by a number of spatially
extended objects which are likely to be \ion{H}{2} regions.

We used the F450W and F702W images to
identify and measure the properties of these \ion{H}{2}
regions. The F450W and F702W observations each consisted of 
two 500\,s exposures, spatially offset from each other by a small
amount. To obtain the final image for each filter, 
the two exposures were registered
and then combined; pixels affected by cosmic-ray events 
were removed in the process.
We decided to transform the {\it HST\/} magnitudes derived from the F450W and 
F702W images to Johnson-Cousins $R$ magnitudes and $B-R$ color 
(Dolphin 2000), so that our results could be compared more easily 
with others in the literature.

The \ion{H}{2} regions were identified using the deeper F702W image.
This was accomplished by first subtracting the backgound signal, 
selecting a threshold magnitude of $\mu_R = 25.6$\,mag~arcsec$^{-2}$
for the analysis, and smoothing the image using 3x3~pixel 
averaging. The MIDAS AIP software package was then used to generate a
catalog of spatially extended objects.
Regions smaller than 4~pixels (with areas $<0.04$\,arcsec$^{-2}$) were
subsequently rejected from the catalog.  We then inspected the
original image by eye to verify that the identified regions had the
location and appearance of \ion{H}{2} regions in SBS~1543+593.
Figure~2 shows the final F702W image and an AIP-derived outer isophote
for the galaxy corresponding to an {\it HST} magnitude 
$\mu_R = 26.8$\,mag~arcsec$^{-2}$.  
Identified \ion{H}{2} regions are labeled
in Figure~2 and listed in Table~1 in order
of increasing ``y'' pixel position.  
Our final catalog comprises 33
likely \ion{H}{2} regions. Note that the bright region observed
spectroscopically by Reimers \& Hagen (1998) corresponds to our
region \#5, but these authors may have also included flux from region
\#7 in their spectrum.  The mask derived from the F702W image was
then applied to the F450W image for the corresponding analysis.

For each of the regions, Table~1 gives an
identifying number, WF3 x-y pixel coordinates, RA and Dec coordinates
derived from the IRAF/STSDAS package METRIC, the $R$ magnitude and $B-R$
color, errors, and parameters describing the sizes and shapes of the
\ion{H}{2} regions (area, semi-major axis, minor-to-major axis ratio,
and position angle).  Table~1 should be
considered a list of {\it candidate} \ion{H}{2} regions, because we
cannot definitively discern the objects' true nature (i.e., whether they are
background galaxies, star clusters, or \ion{H}{2} regions) based on
magnitude and color information alone. On the other hand, most of them
are likely to be \HII\ regions; certainly this is the case for all
the objects covered by our spectroscopic observations described below.

We also determined the total $R$ magnitude of the galaxy (within the
26.8 mag arcsec$^{-2}$ isophote shown on Figure~2) after subtracting
the QSO and the star. 
We found R$_{tot} = 16.22 \pm 0.20$  and $(B-R)_{\rm tot} = 0.20  \pm  0.30$. 
The errors quoted here are total errors and
include the photometric and background errors,
the errors from
the subtraction of the quasar and foreground star, 
errors for transformation to Johnson-Cousins
$B$ and $R$, and zero point errors. 
The total magnitude error is dominated by the errors 
in the subtraction of the QSO and the star. Of course
the magnitudes quoted also depend on the surface brightness 
level to which they are measured.
Thus, our total $R$-band magnitude is significantly brighter than 
that determined by Reimers \& Hagen (1998), who measured $R = 16.9$ 
to the 24\,mag~arcsec$^{-2}$ isophote, but is in good agreement 
with the value of $R = 16.3$ derived by Bowen et al. (2001a) 
from their fit to the surface-brightness profile.  

From Schlegel, Finkbeiner, \&
Davis (1998) we find that the Galactic foreground extinctions are
relatively small, $A_B = 0.067$ and $A_R = 0.042$. 

\subsection{Spectroscopy}

Spectroscopic observations 
of the QSO and some \ion{H}{2} regions in SBS~1543+593 
were obtained in service mode 
with the double-beam Cassegrain spectrograph (ISIS)
of the 4.2\,m William Herschel Telescope (WHT) 
on La Palma, Canary Islands,
and with the Blue Channel spectrograph of the 
6.5\,m Multiple Mirror Telescope (MMT). 
In all cases we used a 2$\arcsec$\ wide slit; other
details of the observations are collected in Table~2.

Figure~2 shows the location of the slit overlaid on the
contour map of the galaxy. 
We attempted to maintain this same slit
location for all spectroscopic observations and were
aided in this by the presence of the QSO. However, we have
no way of estimating uncertainties in the emission line fluxes
which may result from small differences in slit positioning
and from seeing fluctuations.
The slit was oriented at position angle PA\,=\,10.5\,degrees 
east of north, so as to include 
the QSO and the brightest \ion{H}{2} region in SBS~1543+593
(region \#5). This precluded us from observing at the
parallactic angle. As can be seen from Figure 2, with this orientation
we also covered four other \HII\ regions, in two close pairs 
(\#12, \#15, and \#27, \#28 respectively); however, each member of the
pair is not resolved. 

In Figure~3 we have reproduced a
portion of the two-dimensional ISIS red spectrum covering the
H$\alpha$, [N~II]~$\lambda 6583$, and [S~II]~$\lambda\lambda 6717,6731$
emission lines. \HII\  regions and the QSO are marked on the $y$-axis.
Interestingly, in addition to the \HII\ regions targeted, 
we also detect emission lines
from a previously unrecognized source, 
labeled ``X'' in Figure~3,  
2.9$\arcsec$\ to the north of region \#5. There
is no evidence of this \ion{H}{2} region in the {\it HST\/} F702W 
image (see Figure~1), nor in the unfiltered STIS image obtained 
by Bowen et al. (2001a).

The spectra were reduced using standard techniques in IRAF; this
included bias substraction, flat fielding, sky subtraction,
and wavelength calibration.
The acquisition of the ISIS red spectrum was accompanied by
observations of two standard stars, G138-31 and Kopff~27.
Both stars were observed through a 6$\arcsec$ wide slit,
under the same seeing conditions as the object, 1$\arcsec$ FWHM, 
and at low airmass (1.06). We chose G138-31 for the flux
calibration. However, in order to investigate systematic
errors due to the flux calibration, we compared
the fluxes derived, both for the QSO and \ion{H}{2} region \#5, 
using the other standard star available, Kopff~27.
We found that the QSO flux is 15\% lower at the blue end 
when calibrated with Kopff~27, and that this offset reduces 
to 4\% at the red end of the spectrum. Thus, we estimate
the systematic flux error
at H$\alpha$ for \HII\ region \#5 to be $\pm$5\%.
One dimensional extractions 
of the \HII\ region spectra are reproduced in Figure~4. 

Figure 5 shows the ISIS spectrum of \HII\ region \#5 obtained with the 
blue arm of ISIS. It includes the H$\beta$ line and the 
[\OIII]~$\lambda\lambda 4959, 5007$ doublet
(\HII\ region \#5 is the only one in which these lines are
detected in our blue spectra). Data obtained over two nights
were combined to derive the final calibrated spectrum. 
The individual spectra were calibrated
relative to the standard star BD~+17~4708, which was observed
on both nights under similar seeing conditions, 
through a 2$\arcsec$ wide slit, and at low airmass (1.0). 
Note, however, that the observations of SBS~1543+593 were obtained at
higher airmasses, 1.5 and 2.0 on the two nights respectively
(see Table 2). Under these
conditions we expect significant atmospheric
dispersion, but the effect is likely to be negligible over the small
wavelength interval between the emission lines of interest, i.e.,
H$\beta$ and [\OIII]~$\lambda5007$.
The temperature sensitive [\OIII]~$\lambda4363$ line remains
undetected in \HII\ region \#5, despite the long
exposure time of 10\,000\,s (see Figure~5). 

Any systematic offset in the flux calibrations of the blue and
red spectra will affect the measured ratio of H$\alpha$ to H$\beta$, 
needed to estimate the internal reddening of \HII\ region \#5, 
and the determination of the nitrogen-to-oxygen ratio. 
Unfortunately, there is no wavelength
overlap between the ISIS red spectrum obtained on April 25, 2000 and
the ISIS blue spectra recorded on June 12 \& 13, 2001. 
We tried to interpolate across this wavelength gap using 
the spectral energy distribution of the QSO, but its rich iron
spectrum makes judging the location of the continuum difficult. 
Thus, we are unable to provide a quantitative assessment of 
systematic flux errors; our best guess is that the offset
between the red and the blue spectra is no larger than 15\%.  

The ISIS blue spectra do not extend to sufficiently short 
wavelengths to include the [\OII]~$\lambda 3727$ emission line.
For this reason, we obtained an additional short exposure
with the Blue Channel of the MMT spectrograph;
the one-dimensional spectrum is shown in Figure~6. 
The conditions were not photometric and we did not
obtain a standard-star observation during this night. However,
there is substantial overlap in the spectral coverage of
the MMT and the WHT spectra. We calibrated the MMT  spectrum with 
reference to the WHT one, by forcing the 
fluxes of the emission lines in common (H$\beta$ and [\OIII])
to agree between the two sets of observations (the observed
mean ratio for the three emission lines in common was
$1.65\pm0.04$).

The rest frame fluxes of all the emission lines detected
are collected in Table~3. The flux errors quoted
are random errors only. They reflect the $1\sigma$ error in the
continuum underlying the line; a $4\sigma$ upper limit is given 
for lines that were not detected.

In addition, the spectroscopy enabled us to make an independent
determination of the redshift of SBS~1543+593.  The observed
wavelengths of the emission lines are listed in Table~3.  We
derive a mean redshift of $z=0.0096 \pm 0.0001$ for SBS~1543+593, which
is the same as the value measured 
by Bowen et al. (2001b) from 21\,cm emission.

\section{Results}

\subsection{Properties of the \ion{H}{2} Regions of SBS~1543+539}

The \ion{H}{2} regions are tracers of the young stellar population
in SBS~1543+539. They are scattered throughout 
the disk of SBS~1543+539. The stars in the disk are not resolved. 
The yellow-reddish color of the unresolved disk light 
indicates that an older and more contiguous stellar population 
underlies the sprinkling of \ion{H}{2} regions (see Figure~1). 
The light of the unresolved stars, 
together with the distribution of \ion{H}{2} regions, 
trace a spiral arm-like pattern.
We therefore classify this galaxy as a type Sm.
According to the definitions by de Vaucouleurs (1959) and
Sandage \& Bingelli (1984), 
Sm galaxies are the latest-type dwarfs in which some
fragmentary traces of a spiral pattern are still present, 
while dwarfs of even later type are classified as Im. 
The latter are galaxies that have a totally chaotic appearance, 
but the differences are really quite subtle. 
(Panel 21 in Sandage \& Bingelli 1984 also serves 
to illustrate the appearance of high surface brightness BCDs, such as I~Zw~18; 
a comparison with Panels 12--17 shows clearly how different BCDs are
from Sm/Im galaxies).

With the {\it caveat} that not all of the candidates listed in 
Table~1 may turn out to be \HII\ regions, it is instructive
to consider their global properties, for comparison with those 
of other nearby galaxies (\S4.2).
A color-magnitude plot of the candidate \HII\ regions 
listed in Table~1 indicates that there are four objects 
with colors or magnitudes which lie outside the main distribution.
It is possible that these extended objects were erroneously 
considered to be candidate \HII\ regions in SBS~1543+539. 
In what follows, we assume a value of $33 \pm 2$ for the total number of 
\HII\ regions in SBS~1543+539.

Following equation (1) of Kennicutt, Edgar \& Hodge (1989),
the $R$-band luminosity function (LF) of the  candidate \HII\ regions
was fitted by a power-law with slope $a = +0.36$. 
The luminosity function is complete down to $R \simeq 23.8$. 
If we interpret the $R$-magnitude as an 
H$\alpha$ flux (by simply scaling the $R$-magnitude 
by a factor of $-2.5$), then the slope becomes $-0.9$. 
The brightest \ion{H}{2} region, \#5, has an H$\alpha$ luminosity of 
$1.2\,h^{-2}_{70} \times 10^{38}$\,ergs~s$^{-1}$ (\S3.2).
This is the luminosity of quite an ordinary \ion{H}{2} region; it is,
for example, a factor of $\sim 40$ lower than the H$\alpha$ luminosity 
of a giant \HII\ region like 30~Doradus in the LMC. \ion{H}{2} region \#5, is also
the most extended one in SBS~1543+539 (see Table~1), with a diameter of 1.9\,arcsec
which translates to a linear size of about 390\,pc (see \S4.1).
The faintest \HII\ regions for which we believe to be complete 
have luminosities of about $1.0\,h^{-2}_{70} \times 10^{37}$\,ergs~s$^{-1}$.

\subsection{Star Formation Rates}

We can use the H$\alpha$ emission line fluxes from the
\ion{H}{2} regions observed spectroscopically to estimate 
star formation rates, as follows.
The heliocentric velocity of SBS~1543+539,
measured from 21\,cm (Bowen et al. 2001b)
and optical emission lines (\S2.2) is
$v_{\rm helio} = 2868$\,km~s$^{-1}$.
Correcting for the motion of the Sun relative to the
local standard of rest (LSR), and for the rotation of the LSR
about the centre of the galaxy, we obtain a galactocentric
velocity $v_{\rm gal} = 3033$\,km~s$^{-1}$.
Ignoring peculiar velocities giving rise to departures from
the Hubble flow, this corresponds to a distance
$d = 43.3\,h_{70}^{-2}$\,Mpc.
At this distance, the H$\alpha$ emission line fluxes 
listed in Table~3 correspond to the following 
H$\alpha$ luminosities:
$1.2\,h^{-2}_{70} \times 10^{38}$\,ergs~s$^{-1}$ (\ion{H}{2} region \#5); 
$0.18\,h^{-2}_{70} \times 10^{38}$\,ergs~s$^{-1}$ (object X);
$0.15\,h^{-2}_{70} \times 10^{38}$\,ergs~s$^{-1}$ (\ion{H}{2} regions \#12,15);
and
$0.24\,h^{-2}_{70} \times 10^{38}$\,ergs~s$^{-1}$ (\ion{H}{2} regions \#27,28).  

With Kennicutt's (1998) H$\alpha$--SFR calibration

\begin{equation}
{\rm SFR~(M_\odot~yr^{-1}) = 7.9 \times 10^{-42}~{\it L}_{H\alpha} (ergs~s^{-1})}
\end{equation}

\noindent we obtain 
SFR\,=\,(9.1, 1.4, 1.2, 1.9)\,$h^{-2}_{70} \times 10^{-4} M_{\odot}$~yr$^{-1}$ 
for the four regions respectively.

The  total $R$-band luminosity of all 33 \ion{H}{2} regions is a factor
of $\approx 4.4$ times that of \ion{H}{2} regions \#5,12,15,27,28.
Thus, we can obtain an approximate estimate of the total 
SFR in SBS~1543+539 by assuming that the $R$-band to H$\alpha$ 
luminosity ratios are roughly the same for all 33 candidate \ion{H}{2} 
regions. This assumption could be questioned because,
for example, object X is not detected in $R$, but it may be 
approximately valid when all the \HII\ regions are considered together.
With this scaling, we estimate that the total H$\alpha$ luminosity of 
SBS~1543+539 is
$\approx 7.7\,h^{-2}_{70} \times 10^{38}$\,ergs~s$^{-1}$ which
corresponds to a total 
SFR\,$\approx 0.006\,h^{-2}_{70} M_{\odot}$~yr$^{-1}$.

\subsection{Element Abundances}

In this section we use the emission line fluxes 
listed in Table~3 to deduce
the nebular abundances of oxygen and nitrogen in \HII\ region \#5.
Normally, the first step is to correct the line fluxes for effects 
of dust reddening. However, in our case the ratio
$F$(H$\alpha$)/$F$(H$\beta$)\,$= 2.4 \pm 0.5$ is 
{\it smaller} (although consistent within the error)
than the case B recombination value of 2.86 for
the `standard' $n_e = 100$\,cm$^{-3}$ and $T_e = 10^4$\,K
\HII\ region density and temperature (Osterbrock 1989).
Evidently, the reddening of \HII\ region \#5 is small
and we therefore neglect it in the calculation of nebular
abundances. [The foreground Galactic reddening 
of E($B-V$)\,=\,0.016 determined from the maps by
Schlegel, Finkbeiner, \& Davis (1998)
is also inconsequential].

\subsubsection{Oxygen}
Approximate estimates of the oxygen abundance in 
\HII\ regions have traditionally been obtained
from the strong line index 
$R_{23} \equiv {\rm (}F_{5007} + F_{4959} + F_{3727}{\rm )}/F_{\rm H\beta}$
which relates (O/H)
to the relative intensities of 
[O~II]~$\lambda 3727$, [O~III]~$\lambda\lambda 4959, 5007$,
and H$\beta$ (Pagel et al. 1979).
Recent reassessments by Skillman, C{\^ o}t{\' e}, \& Miller (2003)
and Kennicutt, Bresolin, \& Garnett (2003)
have shown that the resulting values of (O/H) are typically accurate
to within $\pm 0.3$\,dex.
In deriving the oxygen abundance
we have made use of the most recent formulation by
Kobulnicky, Kennicutt, \& Pizagno (1999) of the
analytical expressions by McGaugh (1991); 
these formulae express (O/H)
in terms of $R_{23}$ and the ionization index 
$O_{32} \equiv {\rm (}F_{5007} + F_{4959}{\rm )} / F_{3727}$\,.
The results are illustrated in Figure 7 which shows the well-known 
double-valued nature of the relation. 
However, in this case the high value of $R_{23}$ 
($\log R_{23} = 0.90 \pm 0.09$) corresponds to the region of the
diagram where the upper and lower branches of the 
$R_{23}$ relation meet. Taking into account the random
errors in both the  $R_{23}$ and $O_{32}$ indices
(calculated by propagating the errors listed in Table~3),
we obtain values of $12+ \log {\rm (O/H)}$ which range from 7.96 to 8.61,
or $12 + \log {\rm (O/H)} = 8.3 \pm 0.3$\,.
These in turn correspond to oxygen abundances of between
1/6 and 3/4 of solar ([O/H]\,$ = -0.78$ to $-0.13$)
adopting the value $12 + \log {\rm (O/H)}_{\odot} = 8.74$
from the recent reappraisal of solar photospheric abundances
by Holweger (2001).

More recently, Kewley \& Dopita (2002)
have used a combination of stellar population synthesis 
and photoionization models to develop a set of 
ionization parameter and abundance diagnostics which
are also based on the use of the strong optical emission lines.
They claim that because their
techniques solve explicitly for both the ionization parameter 
and the chemical abundance, their diagnostics 
are an improvement over earlier methods, such as $R_{23}$.
Applying their algorithms to our measurements in Table~3,
we obtain $12 + \log {\rm (O/H)} = 8.15^{+0.3}_{-0.1}$,
$\approx 0.15$\,dex lower than our previous estimate.

We can also use the scaling relation between the $N2$ parameter
($N2 \equiv \log ({\rm [N~II]}~\lambda 6583/$H$\alpha$)
and the oxygen abundance recently proposed by 
Denicol\'{o}, Terlevich \& Terlevich (2002) to obtain an
independent estimate of (O/H) from the relative strengths
of [N~II]~$\lambda 6583$ and H$\alpha$.
Their equation

\begin{equation}
{\rm 12+log(O/H)~=~9.12(\pm 0.05)+0.73(\pm 0.10) \times N2}
\end{equation}

\noindent yields $12 + \log {\rm (O/H)} = 8.2 \pm 0.2$. The 
quoted error here includes both the random error in our 
measurement of the $N2$ index and the errors quoted in eq. (2).  
The latter are from a least-squares fit
to data of Denicol\'{o} et al. (2002), 
which extend over an oxygen abundance range
from 1/50 of solar to twice solar, and thus are an
indication of the scatter of the data from the 
best-fit calibration. We place high confidence in our
determination of the oxygen abundance with the $N2$ index, 
because the [N~II]~$\lambda 6583$ and H$\alpha$ lines were 
measured from the same spectrum. 
Moreover, this spectrum was recorded at low airmass (see \S2.2). 
However, we note that the oxygen abundance derived using the $N2$ index 
is in good agreement with the values obtained above using
the ratios of [O~II], [O~III], and H$\beta$, even though
these emission lines 
were measured from different spectra 
obtained under less-than-optimal conditions.

Thus, taking all of the above into account, our best
estimate of the oxygen abundance in \HII\ region \#5
is $12 + \log {\rm (O/H)} =  8.2 \pm 0.2$.

\subsubsection{Nitrogen}

Thurston, Edmunds, \& Henry (1996) have used
grids of photoionization models to develop 
a new algorithm for estimating N abundances
based only on observations of the strong
[\NII], [\OII], and [\OIII] lines.
Unfortunately, their method cannot be applied here
because their photoionization models do not extend to values 
of $\log R_{23} > 0.75$, as found in \HII\ region \#5.
Instead, we followed van Zee et al. (1998) 
in using the value of the oxygen abundance deduced above to estimate
the electron temperature and the corresponding ionization correction.
Specifically, for the large sample of \HII\ regions
studied by van Zee et al. and Kobulnicky, Kennicutt, \& Pizagno (1999),
 an oxygen abundance of $8.2 \pm 0.2$
corresponds to $T_e \approx (11\,300 ^{+2\,100}_{-1\,500})$\,K.
Using the software by Shaw \& Dufour (1995), which
is built on the five-level atom program FIVEL
and implemented in the STSDAS NEBULAR package,
we found that the [\SII]~$\lambda\lambda 6717, 6731$
line ratio implies a density $n_e \approx 80$\,cm$^{-3}$.
With these values of $T_e$ and $n_e$, the task IONIC
then gives the abundance of N$^+$. With ionization corrections
derived from eq. (13)--(15) of Izotov, Thuan, \& Lipovetsky (1994),
we finally obtain 
$\log {\rm (N/O)} = -1.41\,(^{+0.10}_{-0.10})~(^{+0.21}_{-0.25})$
where the first source of error arises from the uncertainty in
$T_e$, while the second is due to the measurement errors
in the fluxes of the H$\alpha$ and [\NII]~$\lambda 6583$ 
emission lines listed in Table~3. 

Adding these two errors in quadrature,
we conclude that $\log {\rm (N/O)}$ 
in \HII\ region \#5 of SBS~1543+539
is between $-1.7$ and $-1.2$, 
or between $\sim 1/8$ and $\sim 2/5$ of the solar 
$\log ({\rm N/O})_{\odot} = -0.81$ (Holweger 2001).

\section{Comparative Analysis of the Properties of SBS~1543+539}

\subsection{Summary of Derived Properties}

We confirm the dwarf galaxy nature of SBS~1543+593.  We find total
magnitudes corrected for Galactic foreground extinction of 
$M_R = -17.0$ and $M_B = -16.8$\,.\footnote{These and all 
subsequent values are for our adopted value of
the Hubble constant $H_0 = 70$\,km~s$^{-1}$~Mpc$^{-1}$;
we have dropped the scaling factor $h_{70}$ for simplicity.}
At a distance $d = 43.3$\,Mpc,
an angular size of 
1\arcsec\ corresponds to 208\,pc; thus the 
area subtended by
the galaxy on the sky, 974\,arcsec$^2$, 
goes together with a physical area of about 
42\,kpc$^2$. Our estimate of the galaxy's 
star formation rate, 
${\rm SFR} \approx 0.006\,M_\odot$~yr$^{-1}$,
then corresponds to a SFR per unit area of 
$\approx 1.4 \times 10^{-4}\,M_\odot$~yr$^{-1}$~kpc$^{-2}$,
and to a gas consumption timescale of $\approx 250$\,Gyr
adopting the \HI\ mass  
$M_{\rm H~I}  = 1.5 \times 10^9 M_{\odot}$
determined by Bowen et al. (2001b). The \HI\ column density seen on the 
sightline to HS~1543+5921 
translates into a (near-)central \HI\ surface density 
of $1.8\,M_{\odot}$~pc$^{-2}$ for SBS~1543+539.
The $M_{\rm H~I}/L_B$ ratio is $\approx 2$ (in solar units).
SBS~1543+593 contains about 33 \HII\ regions. The \HII\ regions' 
H$\alpha$ luminosity function can be fitted with
a power law with slope $-0.9$.
The brightest \HII\ region in SBS~1543+593
has a relatively modest H$\alpha$ luminosity of 
$1.2 \times 10^{38}$\,ergs~s$^{-1}$, only
$\sim 1/40$ that of 30 Doradus in the LMC. It is about 390\,pc in diameter.
In this region we estimate the oxygen abundance to be
$12 + \log ({\rm O/H}) = 8.2 \pm 0.2$ and 
$\log {\rm (N/O)} = -1.40^{+0.2}_{-0.3}$.

\subsection{Comparison with Local Dwarfs}

Star-forming dwarf galaxies exhibit a wide range of central surface 
brightnesses, ranging from galaxies 
brighter than 22~mag arcsec$^{-2}$, usually identified as BCDs,
to low surface brightness galaxies 
which include both dwarf Irregular 
and dwarf Spiral galaxies. 
With an absolute magnitude $M_B = -16.8$,
SBS~1543+539 is certainly a dwarf galaxy, 
$\approx 25$ times fainter than $L^{\ast}$
(adopting $M^{\ast}_{\rm B} - 5 \log h_{70} = -20.23$
from Norberg et al. 2002).
It is, however, near the borderline between 
high- and low-surface brightness galaxies
conventionally taken to be at a
central surface brightness of 23.0~mag~arcsec$^{-2}$
following Bothun, Impey, \& McGaugh (1997).
With the color correction $(B-R) = +0.2 \pm 0.3$
we measure from our {\it HST} images (\S2.1),
the central surface brightness
$\mu_R (0) = 22.6$\,mag~arcsec$^{-2}$ 
reported by Bowen et al. (2001a)
translates to $\mu_B (0) = 22.8 \pm 0.3$\,mag~arcsec$^{-2}$.
Bowen et al. {\it assumed} a redder color,
$(B-R) = +0.78$, deemed to be typical of LSB galaxies,
and thus deduced a central surface brightness fainter than 
Bothun et al.'s criterion. 
The $(B-R)$ color we derive is at the blue end of the 
range observed in local dwarfs (see Schulte-Ladbeck \& Hopp 1998) but,
in any event, it is clear that SBS~1543+539 is 
not a HSB dwarf, nor a BCD. We classify it as a type Sm, since
a slight semblance of a spiral pattern can be discerned in our images.
We now show that its properties are similar to those 
of the local dwarf galaxy population.

The most recent study of the global properties of late-type dwarf
galaxies is that by van Zee (2001). The galaxies which she studied have
central surface brightnesses between 20.5 and 25\,mag~arcsec$^{-2}$;
scale lengths from about 0.2 to 4.3\,kpc;
SFRs between $10^{-5}$ and $10^{-1} M_{\odot}$~yr$^{-1}$;
\HI\ masses between $3 \times 10^6$ and $4 \times 10^9 M_{\odot}$; 
gas depletion time scales between 5 and 4500\,Gyr; 
and $M_{\rm H~I}/L_B$ ratios from 0.1 to 6.6\,. 
SBS~1543+539 is well within
this distribution in terms of all of these properties. 
The \HI\ surface density of SBS~1543+539 is comparable to the average for LSB  
galaxies (de Blok, McGaugh, \& van der Hulst 1996).
As can be appreciated from Figure~8,
the oxygen abundance, $12 + \log ({\rm O/H}) = 8.2 \pm 0.2$,
is consistent with those of other dwarf Irregulars of
similar $M_B$ (see also Hunter \& Hoffman 1999).
 
The H$\alpha$ LF of the \HII\ regions of SBS~1543+539 has a 
slope of $-0.9$.  This compares well with the range of slopes,
from $-0.64$ to $-2.4$, observed by Youngblood \& Hunter (1999) 
in a sample of local Irregular, BCD, and starburst galaxies. 
Furthermore, the LF slope is quite 
typical given the absolute blue magnitude,
SFR per unit area, and luminosity of the brightest \HII\ region of SBS~1543+539 
(compare with Figure~4 of Youngblood \& Hunter). 
In Figure~9 we show  that the total number of \HII\ regions in SBS~1543+539
is also normal, when compared with other dwarf galaxies of similar blue luminosity, 
and that its brightest \HII\ region has the expected H$\alpha$ luminosity.
The first ranked \HII\ regions observed by Youngblood \& Hunter
typically have sizes of a few hundred pc; 
with a physical diameter of 390\,pc, \HII\ region \#5 in SBS~1543+539 
is normal in that respect too.

In summary, on all accounts
SBS~1543+539 appears to be a garden-variety, late-type, 
star-forming, gas-rich dwarf galaxy.

\subsection{Comparison with DLAs}

On the basis of the limited information available,
SBS~1543+539 also seems to be a relatively normal low-redshift
DLA. Its \HI\ column density, $\log N$(\HI)$= 20.35$,
is near the low end of the distribution
[Rao \& Turnshek (2000), for example, cataloged 
DLAs with $\log N$(\HI) in the range 20.30 to 21.63]
but, given the steep slope of the power-law distribution of values of $N$(\HI), 
low column density DLAs are more common than high column density ones.
It is, however, one of the least luminous among the galaxies
which are either candidate or confirmed hosts of DLAs
at $z < 1$ (Boissier et al. 2003; Rao et al. 2003).
On the other hand, its metallicity of $\approx 1/3$ solar, 
as deduced from the oxygen abundance, 
is near the upper end of the range for DLAs
at $z < 1$  which stretches from solar to
$\approx 1/20$ of solar (Pettini et al. 1999;  
Kulkarni \& Fall 2002).
This of course assumes that absorption line measurements
of the DLA in the spectrum of the QSO 
HS~1543+5921 will yield abundances in the neutral ISM
which are in agreement with those of the ionized gas in \HII\ region 
\#5. Whether this is the case will only be established 
with forthcoming {\it HST} observations, as we now discuss.

\section{Pointers for Future Absorption Line Spectroscopy}

The optical data for SBS~1543+593 presented in this study 
will be an important
reference with which the results of ultraviolet absorption
line spectroscopy of HS~1543+5921 will be compared. 
Here we consider some of the
issues which can be addressed by such a comparison.

\subsection{Degree of Metal Enrichment}

As highlighted above, a topic of immediate interest is to establish
whether emission and absorption line analyses yield
concordant measures of the metallicity of the ISM in this
dwarf galaxy. While this is the case for the ISM in the
solar vicinity (see, for example, the discussion of this point
in Kennicutt et al. 2003), there have been suggestions that 
in less chemically evolved galaxies the ISM may not be so
well mixed and, in particular, that \HII\ regions may
exhibit a degree of self-pollution (Kunth \& Sargent 1986;
Aloisi et al. 2003).
The data currently available for SBS~1543+593 are of insufficient
spectral resolution to clarify the situation.
Specifically, the absorption lines measured in the available
STIS spectrum of the QSO by Bowen et al. (2001a) are all strong
and saturated; at the resolution of those data (200--300\,km~s$^{-1}$),
their observed equivalent widths are consistent
with a very wide range of possible values of column density.

We illustrate this point with the case of \OI~$\lambda 1302.17$.
In the STIS spectrum this line is blended with \SiII~$\lambda 1304.37$;
for simplicity here we assume that the two lines are of comparable
strength (the details of the decomposition of the blend are only
a second order effect for the current purpose) and estimate
a rest-frame equivalent width $W_0(\lambda 1302) \approx 0.65$\,\AA\
from the measurement reported by Bowen et al. (2001a).
This value of $W_0$ corresponds to a column density
$N$(\OI)\,$ > 9.0 \times 10^{14}$\,cm$^{-2}$, this
being the lower limit which applies in the optically thin case
(we have adopted the $f$-value of the transition from the
recent compilation by Morton 2003).
The corresponding oxygen abundance, obtained by dividing
by the neutral hydrogen column density
$N$(\HI)\,$= 2.2 \times 10^{20}$~cm$^{-2}$
measured by Bowen et al. (2001a), is 
${\rm (O/H)} > 4.0 \times 10^{-6}$, or $> 1/135$ of solar.
This value, which is 
$\sim 40$ times lower than that of \HII\ region \#5,
is however a strict lower limit. Indeed, for the line
to be optically thin, the velocity dispersion
parameter of the gas, $b$, must be several 
hundred km~s$^{-1}$.\footnote{This and subsequent values
of $b$ are `equivalent' $b$-values assuming that all
the absorbing gas is located in one `cloud'.
More realistically, we expect the absorption lines 
to consist of several components separated in velocity.
In the customary notation, $b = \sqrt{2} \sigma$,
where $\sigma$ is the one-dimensional velocity dispersion
of the absorbers along the line of sight.}
On the other hand, $W_0(\lambda 1302) \approx 0.65$\,\AA\
would be consistent with the \HII\ region abundance of 
$\approx 1/3$ solar if $b$ were $\simeq 32$\,km~s$^{-1}$, a value
which is quite consistent with the 21\,cm emission profile
recorded by Bowen et al. (2001b).
Such a wide range of neutral oxygen column densities,
and corresponding oxygen abundances, is not unusual 
when dealing with 
low resolution spectra of saturated absorption
lines (e.g. Pettini \& Lipman 1995).

In reality, \OI~$\lambda 1302.17$ is seldomly useful
for abundance measurements, even in high resolution
spectra, because the line is nearly always saturated in DLAs.
More appropriate is the weaker \SII~$\lambda\lambda 1250.6, 1253.8, 1259.5$
triplet. Assuming a solar S/O ratio, as found in 
Galactic stars (Nissen et al. 2003) and nearby \HII\ regions
(e.g. Garnett 2003), we predict 
$N$(\SII)\,$ = 1.15 \times 10^{15}$\,cm$^{-2}$ if the
cool, neutral ISM of SBS~1543+593 has the same
metallicity as \HII\ region \#5.
Saturation of the \SII\ lines should be a tractable
problem at this column density, provided $b$ is greater
than $\sim 15$\,km~s$^{-1}$.

\subsection{Nitrogen as a Measure of Ionization Corrections in DLAs}

Measurements of the column density of \NI\ in the spectrum of HS~1543+5921
(from either the $\lambda 1200.0$ or the $\lambda 1134.7$ triplet)
may help resolve a current controversy concerning the abundance
of N in DLAs, which in turn is relevant to understanding the main
sources of N at low metallicities. This is a topic which has generated
some debate recently (e.g. Pettini et al. 2002; Prochaska et al. 2002;
Centuri\'{o}n et al. 2003; Meynet \& Pettini 2003; Molaro et al. 2003).

Clues to the nucleosynthetic origin of N are provided by its abundance
relative to that of O (a product of massive stars which
explode as Type II supernovae) and, to this end, it has become customary 
to examine the behaviour of (N/O) vs. (O/H) as in Figure~10.
Nitrogen has both a primary and a
secondary component, depending on whether the seed carbon and
oxygen from which N is synthesized are those manufactured by the star 
during helium burning,
or were already present when the star first condensed out of the 
interstellar medium.
In \HII\ regions of nearby galaxies, (N/O)
exhibits a strong dependence on (O/H) when the latter is greater
than $\sim 2/5$ solar; this is generally interpreted 
as the regime where secondary N becomes 
dominant. At low metallicities on the other hand, 
when $12 + \log {\rm (O/H)} \simlt 8.0$ 
(that is, $\simlt 1/5$ solar),
N is mostly primary and tracks O; this results in a 
plateau at $\log {\rm (N/O)}~\simeq -1.5$. 
The main sources of primary N are thought to be
intermediate mass stars
($4 \simlt M/M_{\odot} \simlt 7$, e.g. Lattanzio, Forestini, \& Charbonnel 2000) which manufacture it
through a process called `hot bottom burning'
in thermal pulses during the 
asymptotic giant branch phase (Renzini \& Voli 1981).

When we measure the abundances of N and O 
(or S, which is sometimes used as a proxy for O) in metal-poor
DLAs we find quite a different picture, as can be readily
appreciated from Figure~10.  The (N/O) ratio in DLAs exhibits
a range of values which extends from the primary
plateau at $\log {\rm (N/O)}~\simeq -1.5$ down to
one order of magnitude less. The scatter is now well 
established and cannot be attributed entirely to
measurement errors. Qualitatively, this difference
between DLAs at high redshifts on the one hand,
and the most metal-poor \HII\ regions
in the local universe on the other, is consistent with the 
idea of a delayed production of primary N from
intermediate mass stars (relative to the near-instantaneous
release of O by Type II supernovae), as pointed out by
Pettini, Lipman, \& Hunstead (1995).
Quantitatively, however, the situation is more complicated.
The large number of DLAs with values of (N/O) below the 
primary plateau has led to speculation that the 
time delay between the realease of N and O 
may be longer than anticipated---which in turn 
would suggest that stars with masses $M < 4 M_{\odot}$
can also contribute to the nucleosynthesis of N,
perhaps aided by stellar rotation (Meynet \& Pettini 2003).
An alternative proposal has appealed to a truncated 
initial mass function (Prochaska et al. 2002).
Molaro et al. (2003) have drawn attention to the 
possibility that there may be a {\it lower} limit to
(N/O) which could be used to infer the level of 
primary production of N by massive stars.

A dissenting opinion to this general picture was put forward 
by Izotov \& Thuan (1999) who were struck by the narrow
range of values of (N/O) they measured in the most
metal-poor BCDs and proposed that the 
$\log {\rm (N/O)} \simeq -1.5$ plateau 
constitutes a {\it lower} limit set by 
the primary production of N by the same massive stars
which release O when they explode as supernovae.
A corollary of this hypothesis is that the lower
abundances of N measured in many DLAs are spurious;
Izotov, Schaerer, \& Charbonnel (2001) attribute them
to an overionization of \NI\ relative to \OI\ 
(and \SII\ in cases where S is used instead of O).

Herein lies the importance of measuring $N$(\NI)
in the DLA in HS~1543+5921. As can be seen from 
Figure~10, the (N/O) ratio we deduce for \HII\ region 
\#5 in SBS~1543+593 is typical of other \HII\ regions
with an oxygen abundance of $\sim 1/3$ solar.
It falls in the region of the diagram where secondary
production begins to make a discernable contribution 
to the overall N yield and where
time delay effects would have little impact
on the measured (N/O) ratio. 
Thus, if the ratio (N/O)
in the cool, neutral gas of SBS~1543+593 turned 
out to be lower than the value measured in 
\HII\ region \#5 (while the oxygen abundance was the same),
overionization of \NI\ relative to \OI\ would be
the most likely explanation. Such an outcome 
would provide support
for the interpretation of DLA abundances proposed
by Izotov and collaborators.

A similar test has recently been performed 
in I~Zw~18, but with conflicting conclusions.
While Aloisi et al. (2003) deduced a value
of $N$(\NI)/$N$(\OI) which is consistent with
(actually higher by a factor of $\sim 2$ than)
the (N/O) ratio in the \HII\ gas, the estimate by
Lecavalier des Etangs et al. (2003)
is one order of magnitude lower than that of Aloisi
et al. (2003). Remarkably, these two studies used the same
{\it FUSE} data! The problem in this case is partly due to the
difficulty of deriving a reliable column density for \OI.
In any case, I~Zw~18 is not ideal for this test,
because the interstellar absorption lines are detected 
against the background of OB stars {\it within} the galaxy,
so that the line of sight {\it must} include ionized,
as well as neutral, gas. Furthermore, there may be concerns with
interpreting absorption due to the superposition of multiple sight-lines.
The experiment is much cleaner
in the case of the SBS~1543+593/HS~1543+5921 pair,
since the QSO is at much higher redshift than the galaxy
and is a point source. Thus we await with anticipation the results of
high resolution ultraviolet observations of this QSO-galaxy pair.

\section{Conclusions}

The main conclusion from the imaging and spectroscopic
observations of SBS~1543+593 presented here is that
this is an Sm dwarf galaxy of intermediate surface brightness,
with properties which are entirely in line with those of other 
local dwarf galaxies.

We have obtained the first determinations of the oxygen and nitrogen
abundances in a {\it bona fide} DLA galaxy using emission line
diagnostics which refer to the ionized gas, 
rather than the neutral interstellar medium more commonly
probed with QSO absorption line spectroscopy.
We deduce ${\rm (O/H)} \approx 1/3$ solar 
and ${\rm (N/O)} \approx 1/4$ solar;
these values are consistent with other properties
of SBS~1543+593, including its morphology which shows it to be
an evolved disk galaxy with a mix of young and old stars.

The value of these measurements is that they provide a reference
for future UV spectroscopy of the DLA. The \HII\ region to which
they refer is the brightest in SBS~1543+593 and is located only
3.3\,kpc from the QSO sightline---over this distance we do not 
expect significant abundance gradients. Available, low resolution,
spectra of the QSO HS~1543+5921 only allows us to establish a 
lower limit to the oxygen abundance which is not particularly
informative, ${\rm (O/H)} >  1/135$ solar, because the 
\OI~$\lambda 1302$ line is saturated. In reality the abundance
of O in the neutral ISM must be much larger, since
this lower limit corresponds to a velocity dispersion of the gas
of several hundred km~s$^{-1}$ which is excluded by the available
21\,cm emission line profile. We make a prediction for the column
density of \SII, whose UV absorption lines are expected to 
be only mildly saturated and should therefore provide a more precise
measure of the abundance of the 
$\alpha$-capture elements than \OI.
We have also highlighted the 
importance of \NI\ as a test of the ionization corrections which may
apply to DLA abundance determinations. The location of \HII\ region \#5
in a plot of (N/O) vs. (O/H), near the intersection of the primary and
secondary levels of nitrogen production, will make it easier
to distinguish between ionization and time-delay effects as the 
cause of the low N abundances reported in several high redshift DLAs.

All in all, the SBS~1543+539--HS~1543+5921 pair is particularly well-suited
for a long-overdue comparison between absorption and emission line
abundance measurements in DLAs, because HS~1543+5921: (1) shines 
near centrally through the disk of SBS~1543+539 and (2)  is 
a bright background point source, rather than an extended cluster 
of stars within the galaxy. In this paper we have provided the 
basic emission line data for SBS~1543+539 necessary for such a comparison;
we now await with anticipation the results of forthcoming
UV spectroscopic observations of HS~1543+5921 with STIS.

\acknowledgments
Support for program \#8570 was provided by NASA through a grant from the Space Telescope 
Science Institute, which is operated by the Association of Universities for Research in Astronomy, 
Inc., under NASA contract NAS 5-26555. 
We are grateful to the time assignment committee of the WHT for the award of service time,
and the staff astronomers who carried out the observations on our behalf.
We made
extensive use of the NASA/IPAC Extragalactic Database (NED), which is operated by the
Jet Propulsion Laboratory, California Institute of Technology,
under contract with NASA.
This work was completed while RS-L was a visitor at the Max-Planck Insitute f\"{u}r 
Extraterrestrische Physics and at the Institute of Astronomy, Cambridge. 
She thanks both institutions, as well as the
Department of Physics \& Astronomy of the University of Pittsburgh, for
making these visits possible. 
We also thank Evan Skillman for several fruitful discussions.

\clearpage
\begin{center}
\begin{deluxetable}{rccccccrccccr}
\tabletypesize{\tiny} 
\tablewidth{0pc} 
\tablecaption{{Candidate HII regions}\label{t:HII}} 
\tablehead{          
\colhead{Id}    &
\colhead{X}     &  
\colhead{Y}     &      
\colhead{RA$-15:44$}    &
\colhead{DEC$-59:12$}   &
\colhead{$R$}     &
\colhead{$\sigma_R$}     &
\colhead{$B-R$}   &
\colhead{$\sigma_{B-R}$}     &
\colhead{Area}  &    
\colhead{a}     &  
\colhead{b/a}   &
\colhead{PA} \\ [.1ex] 
\colhead{}   &  
\colhead{(pix)} &
\colhead{(pix)} &
\colhead{(J2000.0)}  & 
\colhead{(J2000.0)} &
\colhead{(mag)} &  
\colhead{(mag)} &
\colhead{(mag)} &  
\colhead{(mag)} &
\colhead{(arcsec$^2)$}&
\colhead{(arcsec)}&  
\colhead{} &  
\colhead{(deg)} } 
\startdata 
 1 & 312.5 & 163.8 & 20.20 & 05.2 & 21.99 & 0.04 &  0.25 & 0.11 &  2.28 &  0.84 & 0.94 &  84 \\
 2 & 166.6 & 172.6 & 22.04 & 02.2 & 24.48 & 0.11 &  1.38 & 1.53 &  0.31 &  0.31 & 0.59 &  87 \\
 3 & 156.8 & 191.5 & 22.23 & 03.8 & 21.91 & 0.06 &  0.85 & 0.23 &  2.54 &  0.96 & 0.74 &  86 \\
 4 & 189.0 & 196.9 & 21.85 & 05.1 & 23.59 & 0.15 &  0.18 & 0.26 &  0.69 &  0.74 & 0.39 &  43 \\
 5 & 347.6 & 202.9 & 19.90 & 09.9 & 21.29 & 0.03 & $-0.01$ & 0.15 &  2.74 &  1.28 & 0.47 &  38 \\
 6 & 179.4 & 197.9 & 21.97 & 05.0 & 24.03 & 0.13 &  0.14 & 0.32 &  0.42 &  0.30 & 0.90 &  15 \\
 7 & 362.8 & 203.6 & 19.71 & 10.3 & 22.87 & 0.12 & $-0.29$ & 0.21 &  0.88 &  0.68 & 0.56 &  58 \\
 8 & 375.2 & 212.5 & 19.58 & 11.5 & 23.44 & 0.18 & $-0.54$ & 0.23 &  0.55 &  0.53 & 0.62 &   1 \\
 9 & 241.7 & 291.3 & 21.52 & 15.5 & 22.95 & 0.07 & $-0.91$ & 0.09 &  0.61 &  0.36 & 0.85 &  74 \\
10 & 265.0 & 293.2 & 21.24 & 16.3 & 24.60 & 0.11 &  2.49 & 0.11 &  0.25 &  0.27 & 0.73 &  12 \\
11 & 289.6 & 295.4 & 20.94 & 17.2 & 23.78 & 0.06 & $-0.11$ & 0.19 &  0.33 &  0.29 & 0.96 &  87 \\
12 & 350.3 & 299.6 & 20.20 & 19.2 & 22.81 & 0.14 &  0.37 & 0.44 &  0.62 &  0.73 & 0.40 & $-45$ \\
13 & 492.9 & 304.8 & 18.44 & 23.5 & 22.56 & 0.05 & $-0.14$ & 0.13 &  1.43 &  0.71 & 0.78 & $-11$ \\
14 & 372.5 & 309.2 & 19.95 & 20.7 & 23.03 & 0.12 &  0.61 & 2.07 &  0.47 &  0.39 & 0.82 & -26 \\
15 & 355.5 & 308.9 & 20.16 & 20.2 & 23.74 & 0.32 &  0.43 & 1.26 &  0.27 &  0.30 & 0.79 &  14 \\
16 & 300.0 & 311.0 & 20.86 & 19.0 & 23.74 & 0.11 & $-0.32$ & 0.17 &  0.32 &  0.37 & 0.60 & $-18$ \\
17 & 365.1 & 321.7 & 20.09 & 21.7 & 22.14 & 0.79 &  0.58 & 0.79 &  1.04 &  0.72 & 0.68 &  60 \\
18 & 346.8 & 343.9 & 20.39 & 23.4 & 22.68 & 0.09 &  0.77 & 0.72 &  0.38 &  0.31 & 0.90 &  87 \\
19 & 342.0 & 351.2 & 20.48 & 23.9 & 22.31 & 0.05 &  0.38 & 0.20 &  0.55 &  0.50 & 0.62 &  70 \\
20 & 386.9 & 354.0 & 19.93 & 25.4 & 23.90 & 0.29 &  0.47 & 0.50 &  0.21 &  0.24 & 0.96 &  53 \\
21 & 231.9 & 359.3 & 21.88 & 21.8 & 24.28 & 0.14 & $-0.43$ & 0.55 &  0.27 &  0.25 & 0.83 &  81 \\
22 & 279.9 & 374.2 & 21.33 & 24.5 & 23.97 & 0.08 & $-0.11$ & 0.21 &  0.28 &  0.23 & 0.82 &  90 \\
23 & 380.1 & 380.8 & 20.10 & 27.8 & 23.57 & 0.05 &  0.69 & 0.08 &  0.27 &  0.32 & 0.80 &  $-2$ \\
24 & 228.0 & 393.1 & 22.04 & 24.9 & 23.12 & 0.10 & $-0.16$ & 0.14 &  0.71 &  0.55 & 0.57 &  87 \\
25 & 286.5 & 397.5 & 21.33 & 26.9 & 23.91 & 0.10 & $-0.08$ & 0.23 &  0.32 &  0.26 & 0.91 &  69 \\
26 & 241.3 & 412.0 & 21.95 & 27.1 & 23.38 & 0.11 &  0.55 & 0.25 &  0.66 &  0.56 & 0.57 &   8 \\
27 & 367.9 & 435.8 & 20.45 & 32.7 & 23.95 & 0.48 & $-0.45$ & 0.51 &  0.25 &  0.25 & 0.91 &  64 \\
28 & 361.4 & 437.3 & 20.53 & 32.7 & 23.87 & 0.22 &  0.15 & 0.23 &  0.30 &  0.33 & 0.76 &  86 \\
29 & 398.6 & 440.3 & 20.08 & 34.0 & 23.13 & 0.11 & $-0.33$ & 0.16 &  0.51 &  0.40 & 0.79 &  $-8$ \\
30 & 438.6 & 441.7 & 19.59 & 35.2 & 23.21 & 0.09 & -0.09 & 0.22 &  0.68 &  0.60 & 0.59 &  57 \\
31 & 417.5 & 457.1 & 19.90 & 36.1 & 23.67 & 0.13 &  1.30 & 1.09 &  0.46 &  0.50 & 0.52 &  31 \\
32 & 517.2 & 467.3 & 18.70 & 39.7 & 24.34 & 0.10 &  0.12 & 0.56 &  0.33 &  0.25 & 0.69 & $-31$ \\
33 & 340.4 & 493.1 & 20.99 & 37.5 & 24.52 & 0.21 & $-0.96$ & 0.24 &  0.27 &  0.31 & 0.79 &  87 \\
Gal.&355.8 & 350.2 & 20.30 & 24.2 & 16.22 & 0.02 &  0.20 & 0.05 & 974.2 & 14.72 & 0.65 &  57 \\
QSO &428.9 & 411.7 & 20.29 & 26.4 &\nodata&\nodata&\nodata& \nodata & \nodata & \nodata & \nodata \\
Star&403.9 & 451.1 & 19.88 & 29.8 &\nodata&\nodata&\nodata& \nodata & \nodata & \nodata & \nodata \\
\enddata
\end{deluxetable}
\end{center}

\clearpage

\begin{deluxetable}{clcrccr}
\tabletypesize{\tiny}
\tablewidth{0pc} 
\tablecaption{Journal of Spectroscopic Observations} 
\tablehead{          
\colhead{Observation Date} &
\colhead{Grating/Detector} & 
\colhead{Wavelength (\AA)} &
\multicolumn{1}{c}{Exp. Time} & 
\multicolumn{1}{c}{Slit Width} & 
\multicolumn{1}{c}{Seeing} &
\colhead{Airmass} \\ [.1ex]
\colhead{} &
\colhead{} & 
\colhead{Central, Range} & 
\multicolumn{1}{c}{(s)} &
\multicolumn{1}{c}{($\arcsec$)} &
\multicolumn{1}{c}{($\arcsec$)} &
\colhead{}
}
\startdata 
2000 04 25 &  WHT ISIS R600R/TEK4      & 6749, 6340 -- 7150 & 10\,000 & 2 & 1 & 1.2 \\
2001 06 12 &  WHT ISIS R600B/EEV12     & 4964, 4050 -- 5800 &  6\,000 & 2 & 1.5 & 1.5\\
2001 06 13 &  WHT ISIS R600B/EEV12     & 4964, 4050 -- 5800 &  4\,000 & 2 & 1 & 2.0\\
2002 04 21 &  MMT Blue Channel 800\,grooves~mm$^{-1}$  & 4275, 3125 -- 5425 & 300 & 2 & 1 & 1.2\\
\enddata
\end{deluxetable}

\clearpage

\begin{center}
\begin{deluxetable}{lccccccccc}
\tabletypesize{\tiny}
\tablewidth{0pc}
\tablecaption{Spectroscopic Results} 
\tablehead{          
\colhead{HII region} &
\colhead{[OII]~$\lambda 3727$} & 
\colhead{[OIII]~$\lambda 4363$} & 
\colhead{H$\beta$} &
\colhead{[OIII]~$\lambda 4959$} & 
\colhead{[OIII]~$\lambda 5007$} & 
\colhead{H$\alpha$} &
\colhead{[NII]~$\lambda 6583$} & 
\colhead{[SII]~$\lambda 6717$} & 
\colhead{[SII]~$\lambda 6731$} 
}
\startdata 
 & \multicolumn{9}{c}{Observed Wavelengths (\AA)}\\
\cline{2-10}\\[.1ex]
\#5 & 3763.6 & \nodata  &  4907.6 & 5006.2 & 5054.3 & 6627.2 & 6647.6 & 6782.1 & 6795.9 \\[.1ex]
\cline{1-10}\\[.1ex]
 & \multicolumn{9}{c}{Rest Frame Flux\tablenotemark{a}\ \ ($10^{-16}$ ergs s$^{-1}$cm$^{-2}$)}\\ [.1ex] 
\cline{2-10}\\[.1ex]
\#5 & $6.10\pm0.48$ & $<$1.72 & $2.13\pm0.43$ & $2.75\pm0.43$ & $8.24\pm0.43$ & 
 $5.13\pm0.09$& $0.31\pm0.09$ & $0.61\pm0.09$ & $0.46\pm0.09$ \\ 
\#X & \nodata & \nodata  & \nodata  & \nodata & \nodata & $0.82\pm0.08$ & $0.11\pm0.08$ & 
 $0.28\pm0.08$ & $0.13\pm0.08$ \\ 
\#12,15 & \nodata & \nodata  & \nodata  & \nodata  & \nodata & $0.65\pm0.08$ & $<0.32$ 
& $0.24\pm0.08$ & $0.23\pm0.08$ \\
\#27,28 & \nodata & \nodata  & \nodata  & \nodata  & \nodata & $1.06\pm0.10$ & $<0.40$ & 
 $0.37\pm0.10$ & $0.15\pm0.10$\\ [.1ex]
\enddata
\tablenotetext{a}{Errors are $1\sigma$ and upper limits are given at the $4\sigma$ level of confidence.}
\end{deluxetable}
\end{center}

\clearpage


\begin{figure}
\plotone{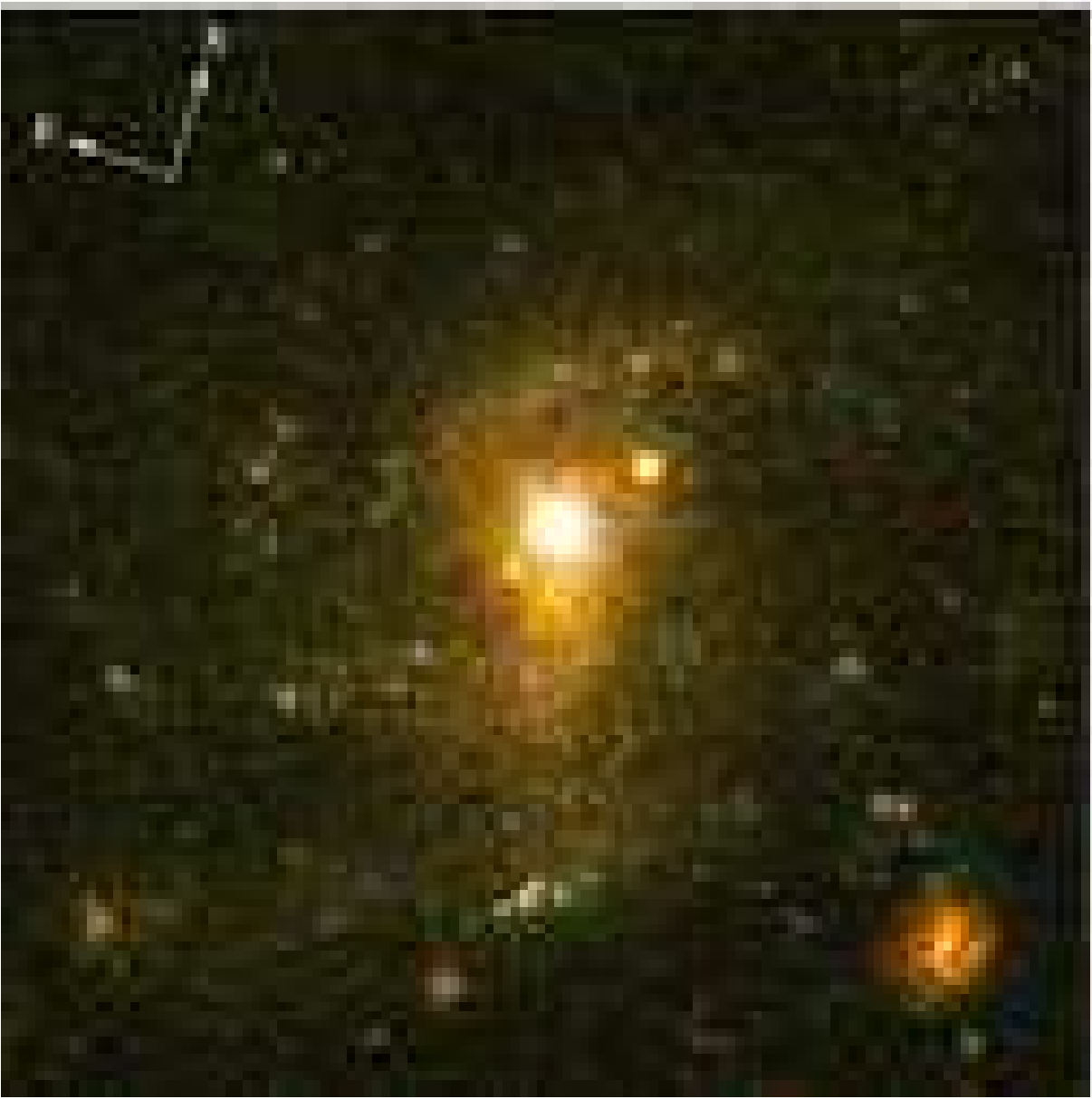}
\vspace{0.5cm}
\caption[SBS1543_col3.eps] 
{
The image of SBS~1543+593 formed by combining all four filters,
as described in the text. The field shown here is a 50$\arcsec$ by 50$\arcsec$
portion of the WF3 chip of WFPC2.
Clearly visible are the bright QSO HS~1543+5921 which
is centrally located relative to the galaxy, a bright foreground
star to the north-west of the QSO, and several background galaxies.
The faint spiral structure of SBS~1543+593 is traced by a number
of spatially extended objects which are likely to be \ion{H}{2} regions.
}
\end{figure}

\clearpage

\begin{figure}
\centerline{\includegraphics[angle=-90, width=20.0cm]{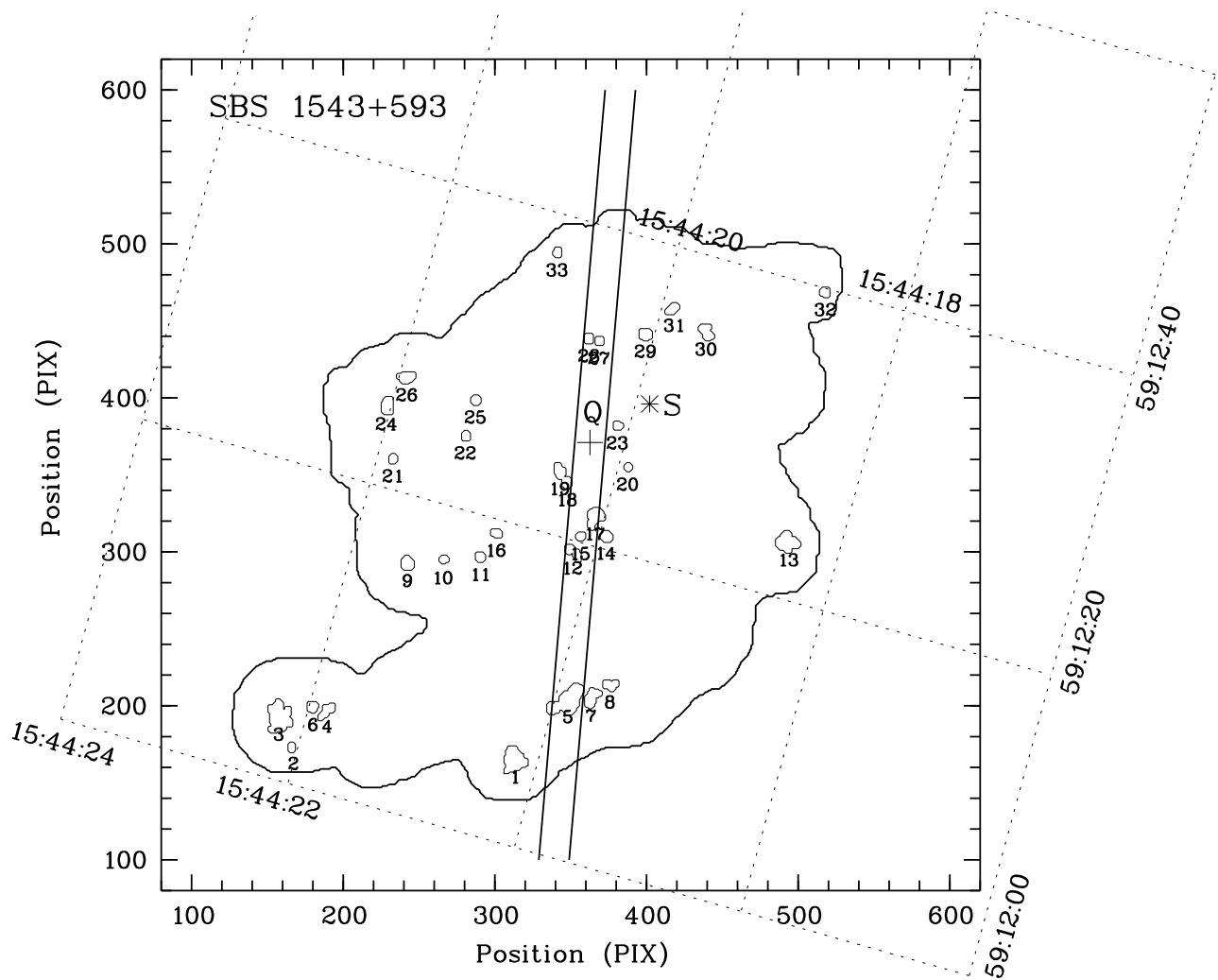}}
 \caption{Map of the F702W flux on WF3. The contour of the entire galaxy is drawn
at $\mu_0(R) = 26.8$~mag~arcsec$^{-2}$, and the isophotes of the compact
regions (candidate \HII\ regions)  correspond to $\mu_0(R) = 25.6$
mag~arcsec$^{-2}$. The cross indicates the position of the QSO; the star
denotes the location of the foreground star. The
position of the slit used for our spectroscopic observations
is illustrated.
}
\end{figure}

\clearpage

\begin{figure}
\plotone{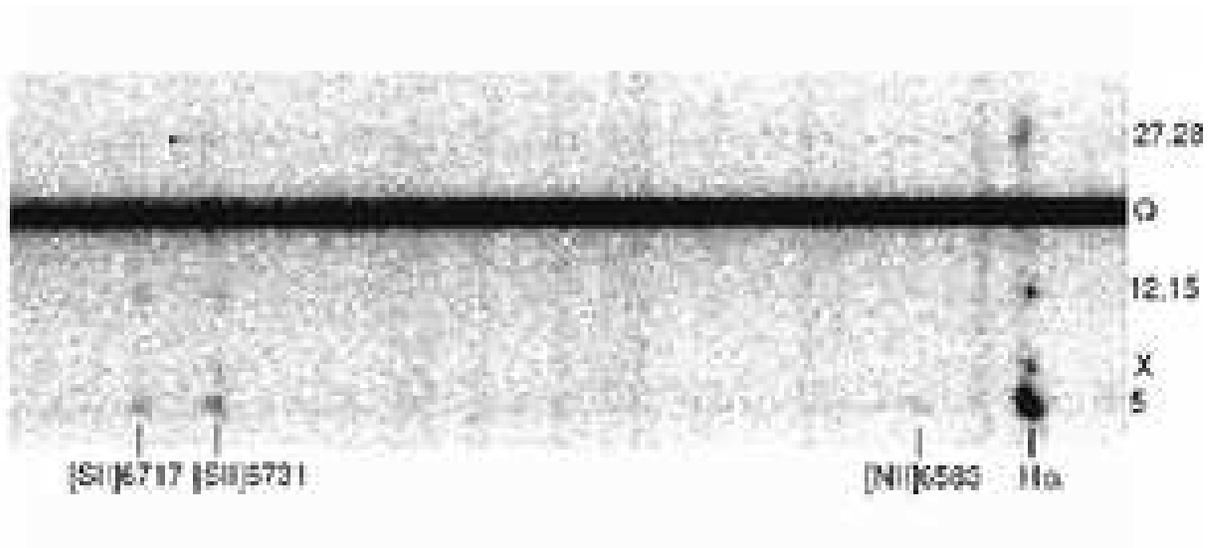}
\caption{Portion of the two-dimensional spectrum of SBS~1543+593 obtained
with the R600R grating of ISIS on the WHT. Emission lines are indicated
below the image (wavelengths increase from right to left); 
\HII\ regions identified in Figure 2 are labeled on the 
right. The \HII\ region marked ``X'' has no visible counterpart in the
{\it HST\/} image---note the lack of a 
detectable continuum signal in this spectrum.
}
\end{figure}

\clearpage

\begin{figure}
\centerline{\includegraphics[angle=-90, width=18.0cm]{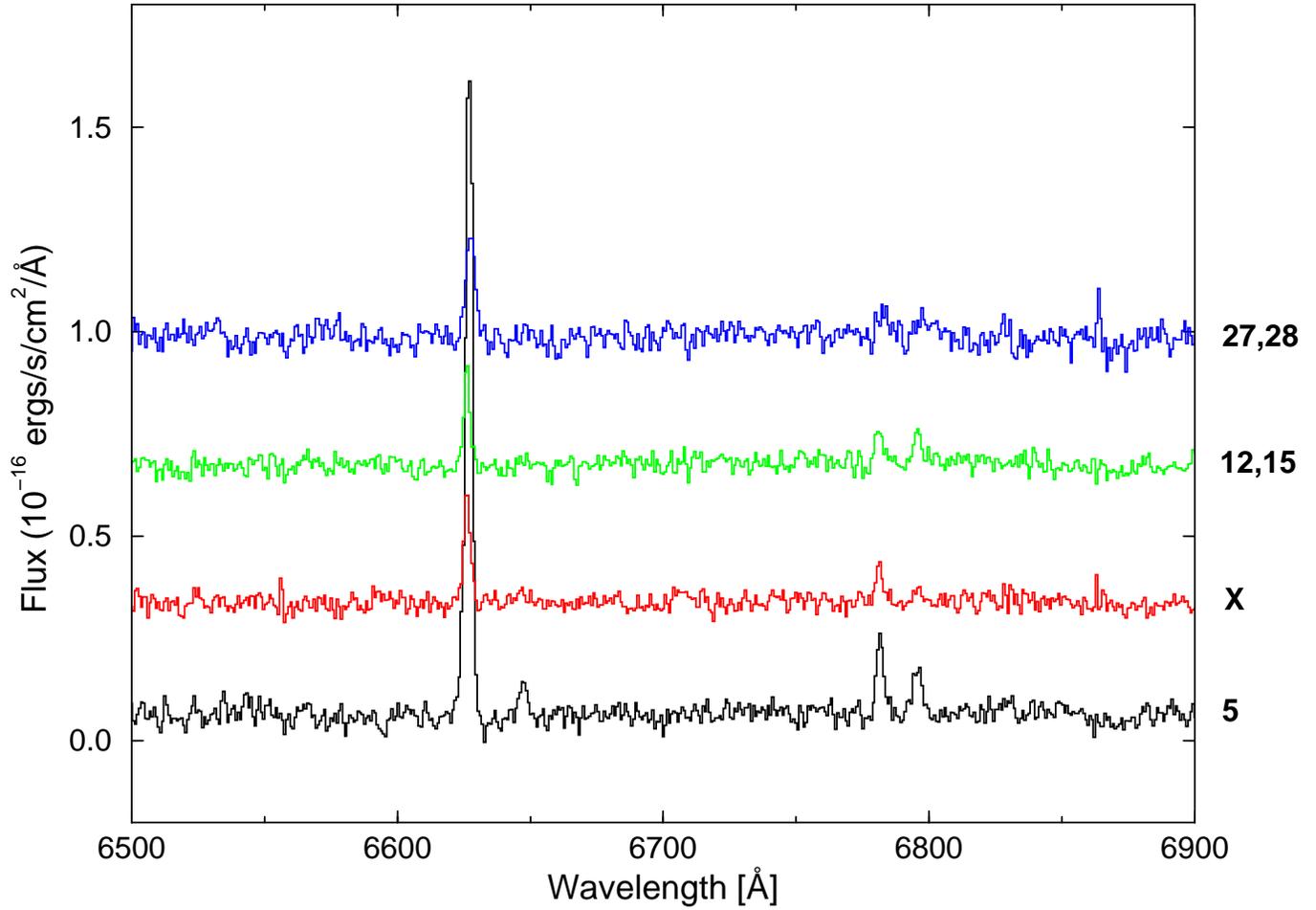}}
\caption{Individual extractions of the \HII\ region spectra shown in Figure~3.
The $y$-axis flux scale refers to the spectrum of \HII\ region \#5;
the others have been offset for clarity.
}
\end{figure}

\clearpage

\begin{figure}
\centerline{\includegraphics[angle=-90, width=18.0cm]{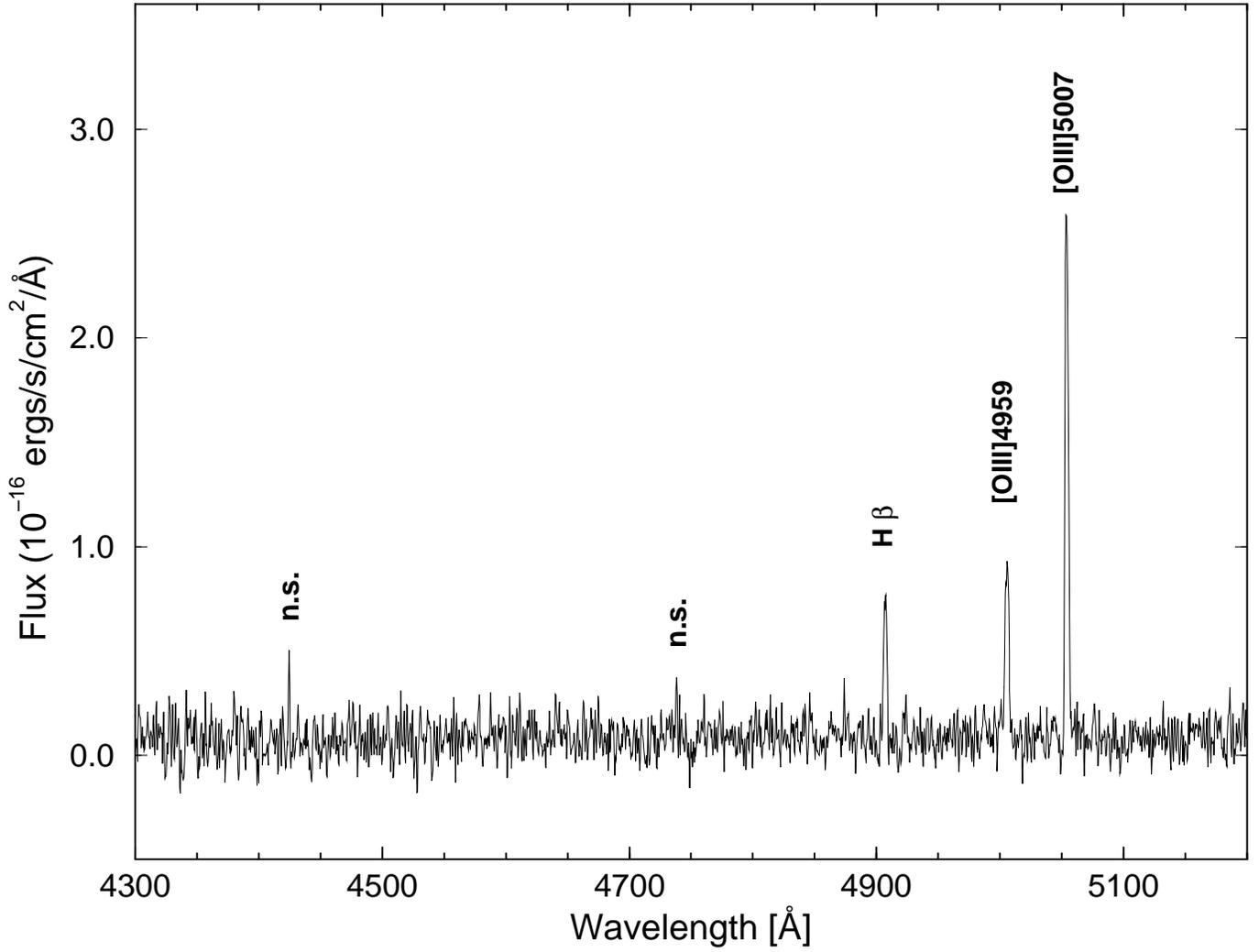}}
\caption{ISIS blue-arm spectrum of \ion{H}{2} region \#5.
No emission lines were detected from the three other \ion{H}{2} regions
shown in Figure 4.}
\end{figure}

\clearpage

\begin{figure}
\centerline{\includegraphics[angle=-90, width=18.0cm]{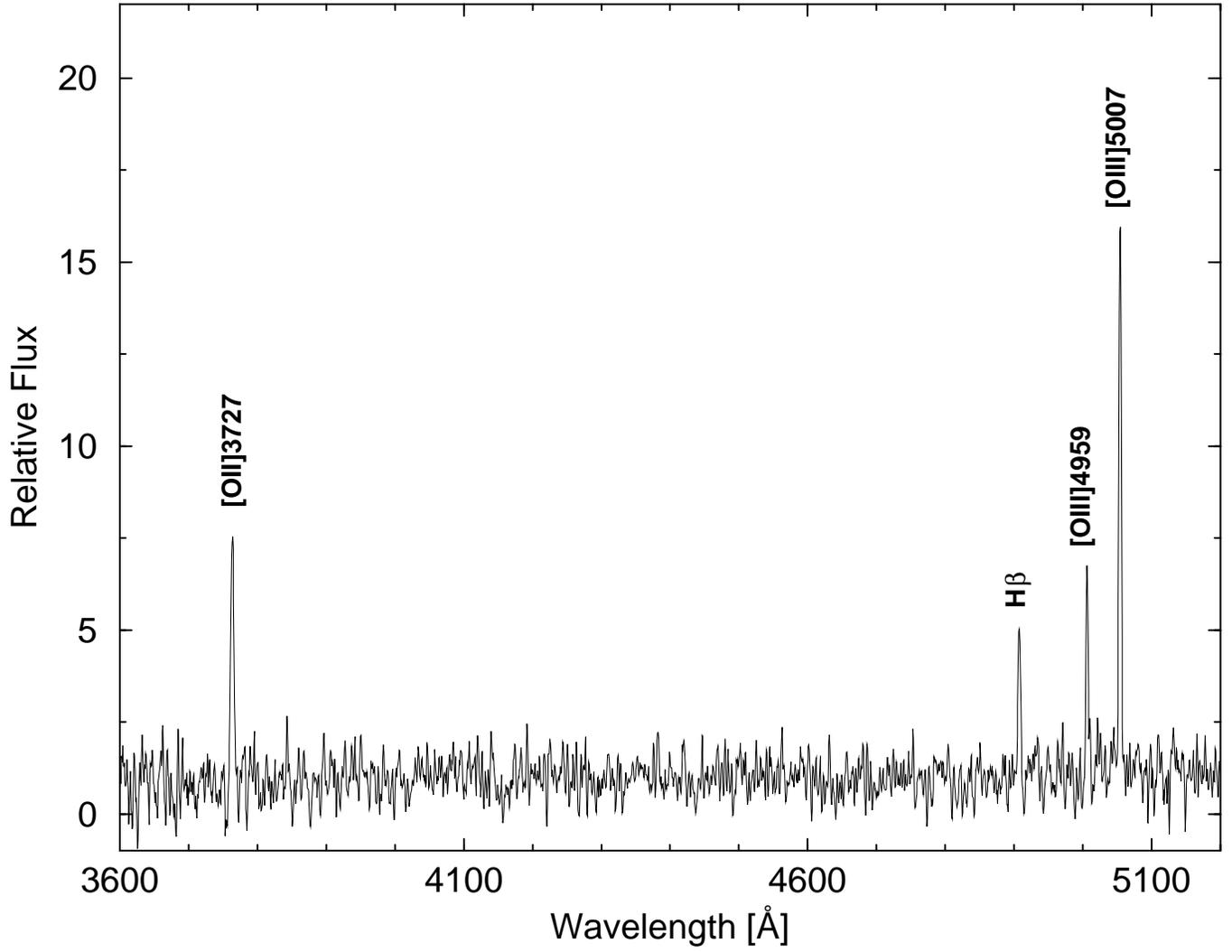}}
\caption{Extracted MMT Blue Channel spectrum of \ion{H}{2} region \#5
including the [OII]\,$\lambda 3727$ line.}
\end{figure}

\clearpage

\begin{figure}
\vspace*{-5cm}
\plotone{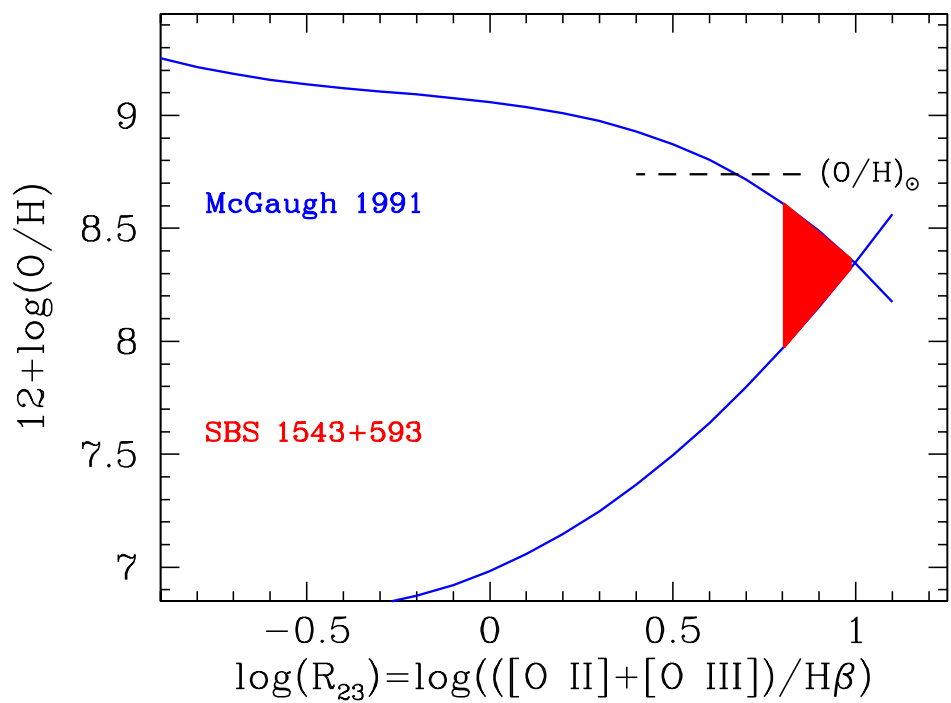}
\vspace{-4.5cm}
\caption{
Oxygen abundance from the $R_{23}$ = ([\OII]+[\OIII])/H$\beta$ ratio.
The continuous lines are the calibration by McGaugh (1991)
for the ionization index $O_{32}$ = [\OIII]/[\OII] appropriate to
SBS~1543+593. The shaded area shows the values allowed
by the measured $R_{23}$ and its statistical $1 \sigma$ error.
The broken horizontal line gives for reference the most recent
estimate of the solar abundance $12 + \log {\rm (O/H)} = 8.74$ (Holweger 2001).}
\end{figure}

\begin{figure}
\vspace*{-5cm}
\centerline{\includegraphics[width=18.0cm]{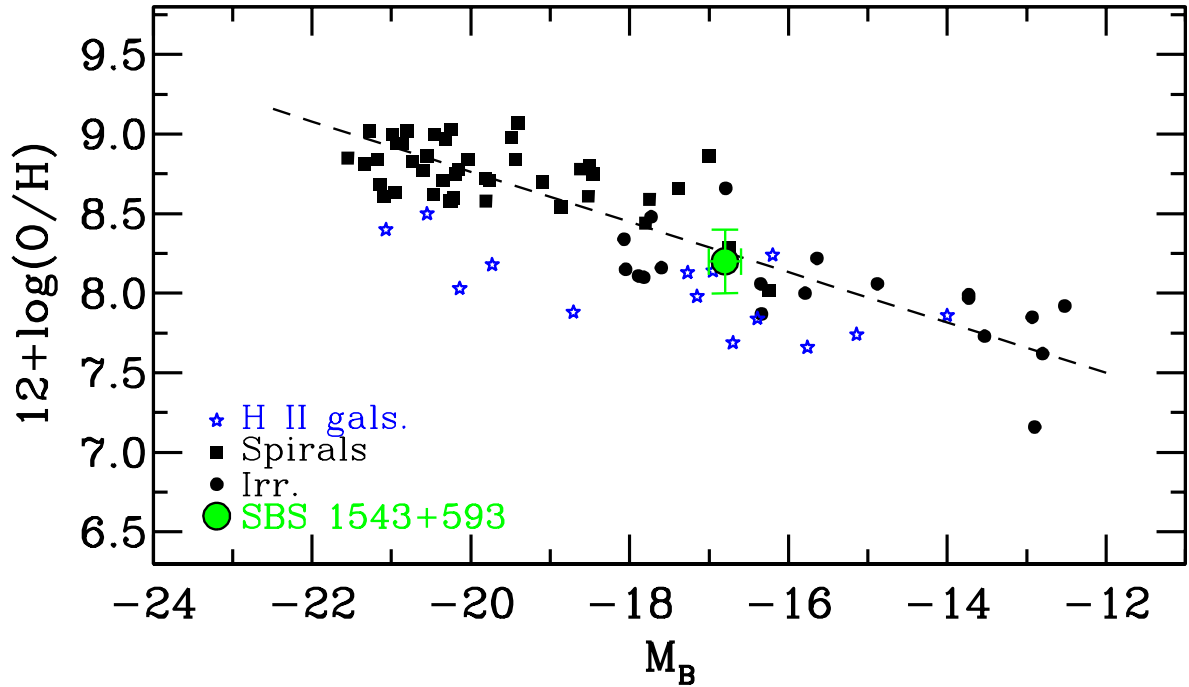}}
\vspace{-2.5cm}
\caption{Metallicity-luminosity relation for local galaxies,
from the compilation by Kobulnicky \& Koo (2000) adjusted to the
$H_0 = 70\,$km~s$^{-1}$~Mpc$^{-1}$ cosmology adopted in this paper.
In the Sun, 12\,+\,log(O/H) = 8.74 (Holweger 2001).
The oxygen abundance $12 +\log {\rm (O/H)} = 8.2 \pm 0.2$
we derive for SBS~1543+593 is typical of its luminosity
$M_B = -16.8 \pm 0.2$\,. 
}
\end{figure}

\begin{figure}
\centerline{\includegraphics[width=18.0cm]{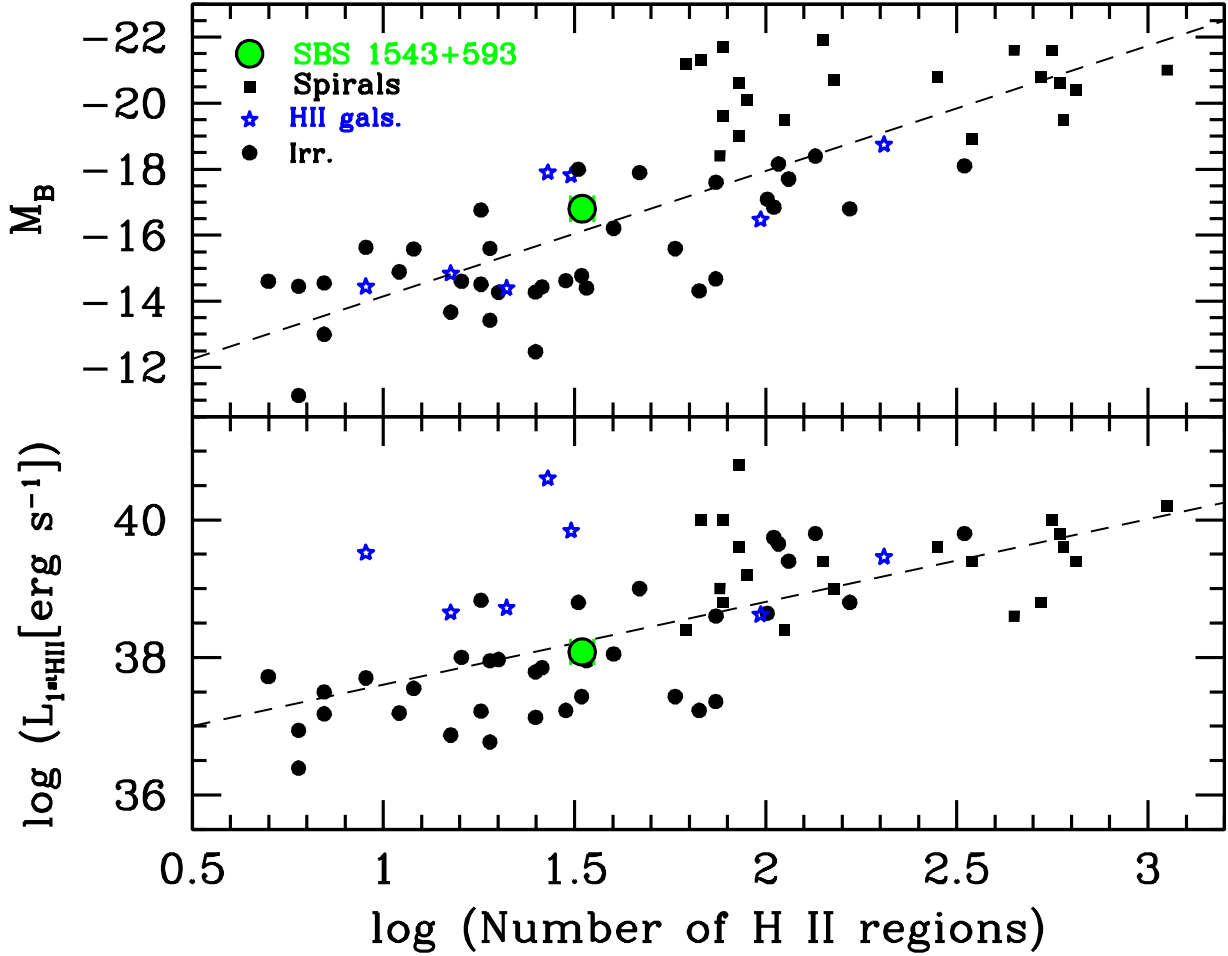}}
\vspace{-1.95cm}
\caption{Relationship between the total number of \HII\ regions in local galaxies
and the blue luminosity of the parent galaxy $\it(Top)$, and the H$\alpha$ luminosity 
of its most luminous \HII\ region $\it(Bottom)$. The points for Irregulars and
\HII\ galaxies (including BCDs and starbursts) from Youngblood \& Hunter (1999)
were adjusted to the $H_0 = 70\,$km~s$^{-1}$~Mpc$^{-1}$ cosmology adopted in this paper.
Additional data for Irregulars and Spirals were taken from Kennicutt, Edgar \& Hodge (1989).
The total number of \HII\ regions we measure in SBS~1543+593 is seen to be typical
of its luminosity. The luminosity of its first ranked \HII\ region, \#5,
agrees with what is seen in other local dwarfs.
}
\end{figure}

\clearpage

\clearpage

\begin{figure}
\vspace*{-2cm}
\centerline{\includegraphics[width=18.0cm]{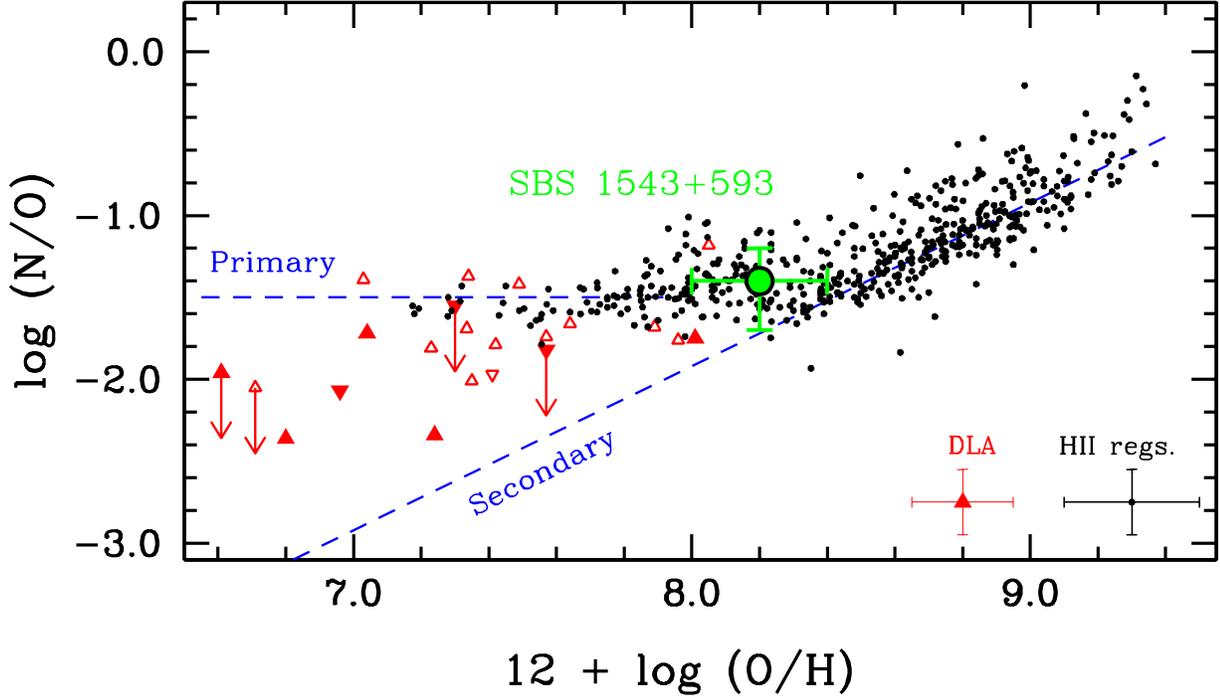}}
\vspace{-2.5cm}
\caption{
Abundances of N and O in extragalactic \HII\ regions
(small dots), \HII\ region \#5 in SBS~1543+593 (large dot with error bars),
and damped Ly$\alpha$ systems (triangles).
Filled 
triangles denote DLAs where the abundance of O 
could be measured directly, while open triangles
are cases where S was used as a proxy for O.
Sub-DLAs (absorption systems with 
$N$(\HI)\,$< 2 \times 10^{20}$\,cm$^{-2}$)
are shown as inverted triangles.
This plot is a recent update of that published 
by Pettini et al. (2002) where references to the original
data can be found. In addition, we have included recent
measurements in DLAs by
Lopez et al. (2002); Lopez \& Ellison (2003); 
and Centuri\'{o}n et al. (2003).
The error bars in the bottom right-hand corner
give an indication of the typical uncertainties.
The dashed lines are approximate
representations of the primary and secondary 
levels of N production (see text).
SBS~1543+593 shows a typical (N/O) ratio
for its oxygen abundance.
In the Sun, $12 + \log{\rm (O/H)} = 8.74$
and $\log {\rm (N/O)} = -0.81$ (Holweger 2001).
}
\end{figure}

\end{document}